\documentclass[aip,jcp,reprint]{revtex4-1} 
\usepackage{graphicx}

\usepackage{amsmath}    
\usepackage{graphicx}   
\usepackage{color}
\usepackage{ulem}
\usepackage{caption}
\usepackage{subcaption}

\usepackage{array}
\usepackage{booktabs}
\usepackage{multirow}

\usepackage{booktabs}

\begin{document}
\title{Kramers turnover: from energy diffusion to spatial
    diffusion using metadynamics}
\author{Pratyush Tiwary}
\email{pt2399@columbia.edu} \affiliation{Department of Chemistry,
  Columbia University, New York 10027, USA.}
	
	\author{B. J. Berne}
	 \email{pt2399@columbia.edu}   
	 \affiliation{Department of Chemistry, Columbia University, New York 10027, USA.}

	\date{\today}
	
	\begin{abstract}
We consider the rate of transition for a particle between two metastable states coupled to a thermal environment for various magnitudes of the coupling strength, using the recently proposed infrequent metadynamics approach (Tiwary and Parrinello, Phys. Rev. Lett. \textbf{111}, 230602 (2013)). We are interested in understanding how this approach for obtaining rate constants performs as the dynamics regime changes from energy diffusion to spatial diffusion.  Reassuringly, we find that the approach works remarkably well for various coupling strengths in the strong coupling regime, and to some extent even in the weak coupling regime.
\end{abstract}

	\maketitle
	\section{Introduction}
It is well known that the dynamics of activated barrier crossing shows
strong sensitivity to the strength of coupling to the environment
\cite{straubberne_review}. Specifically, for a prototypical setup with
two stable states separated by a barrier, the rate constant $k$ for
barrier crossing first increases, and then decreases, as the coupling
to the environment increases.  This non-monotonic behavior of the rate
constant, known as Kramers' turnover, is of fundamental interest in
chemical dynamics, and manifests itself in a range of practical
scenarios, including but not limited to isomerization reactions,
protein folding and even
excitation energy transfer in light-harvesting
systems\cite{thirumalai_kramers,besthummer_kramers,silbeycao_prl}. It has been thoroughly investigated
through numerical, analytical and experimental studies over the past
few decades
\cite{pollak_kramers,straubberne_review,hynes_arpc1985,skinner1980,haynes1994theory}. It is well
accepted that in the low coupling regime the rate constant is small
due to poor exchange of energy with the environment. Either the system
rarely gains enough energy to cross the barrier, or when it does so,
it is unable to dissipate this energy and settle in the product
state. On the other end, in the high coupling regime, spatial
diffusion across the barrier top becomes the rate limiting factor, and
increasing the environmental coupling leads to decrease in the rate
constant.

In principle one can use molecular dynamics (MD) simulations to
directly measure the rate constants for activated barrier crossing
without making any assumptions on the nature of the dynamics regime. Here
this coupling is generally implemented through a friction coefficient
that quantifies the rate of collisions between the system of interest
and a thermal bath. This however becomes a challenging task when the
barrier height is much larger than the thermal energy $k_B T$, and it
becomes difficult to observe sufficiently many or any barrier crossing
events in MD given computer time limitations. To deal with this debilitating problem, over the years several enhanced sampling schemes have been proposed that accelerate
barrier crossing events in a controllable manner. While many of these
enhanced sampling methods concern recovering the underlying free
energy landscape, some are designed to calculate the actual rate of
barrier crossing \cite{reacflux,grubmuller,Voter-PRL-1997,throwingropes,sisyphus2,meta_time}.

In this short communication, we consider one such recently proposed
method, the so-called `infrequent metadynamics' approach
\cite{meta_time,pvalue}, which has recently been applied successfully
to obtain rate constants in a variety of complex molecular systems
\cite{trypsin,fullerene}, and is briefly summarized in
Sec. \ref{metad}. Given the potential benefit of this approach and its
increasing popularity, in this work we explore its robustness in
obtaining rate constants for a model 2-state system \cite{deleonberne}
with respect to varying coupling strength to a thermal bath
implemented via Langevin dynamics \cite{bussi_langevin}. Through a
large number of independent simulations, we identify the environmental coupling
regime in which the dynamics from infrequent metadynamics is correct,
by comparing against much longer unbiased MD runs. We find that 
infrequent metadynamics gives correct rates across
several orders of magnitude variation in the environmental coupling,
as long as one stays in the high coupling regime. It also reproduces
the pronounced change in rate associated with Kramers' turnover. It
tends to become less reliable as the coupling constant is made very
low and the system enters the deeply underdamped regime. Thankfully,
most biomolecular systems, which are the target systems for the
method \cite{trypsin,fullerene}, involve large numbers of solvent atoms executing rapid thermal
motions where one expects the moderate to high friction regime to be
applicable. As we find in this work, the infrequent metadynamics
approach is indeed reliable in this regime.

\section{Dynamics from infrequent metadynamics}
\label{metad}

Metadynamics is an enhanced sampling approach that begins by
identifying a small number of slowly changing order parameters, called
collective variables (CVs) \cite{tiwary_rewt,wtm}. By periodically
adding repulsive bias in the regions of CV space as they are visited,
the system is encouraged to escape stable free energy minima where it
would normally be trapped. The traditional objective of a metadynamics
run is to recover the underlying true free energy surface as a
function of the deposited bias \cite{tiwary_rewt}. Recently, Tiwary
and co-workers extended the scope of metadynamics by showing how to
extract unbiased rates from biased ones with minimal extra
computational burden\cite{meta_time,pvalue}, inspired by works such as Ref. \onlinecite{grubmuller,Voter-PRL-1997}. The central idea was to
deposit bias infrequently enough compared to the time spent in the
transition state regions. Through this `infrequent metadynamics'
approach one increases the likelihood of not corrupting the transition
states through the course of metadynamics, thus preserving the
sequence of transitions between stable states. One can then access the
acceleration of transition rates achieved through biasing by appealing
to generalized transition state theory\cite{straubberne_review} and
calculating the following simple running average\cite{meta_time}:
\begin{equation}
\alpha = \langle e^{\beta V(s,t)} \rangle _t
\label{acceleration}
\end{equation}
where $s$ is the collective variable being biased, $\beta = {1 \over k_B T}$ is the
inverse temperature, $V(s,t)$ is the bias experienced at time $t$ and
the subscript $t$ indicates averaging under the time-dependent
potential.

This approach assumes that one has the correct slow order parameters
or CVs that can demarcate all relevant stable states of
interest. Whether this is the case or not can be verified \textit{a
  posteriori} by checking if the transition time statistics conforms
to a Poisson distribution\cite{pvalue}. The current work assumes one
has the right order parameters or collective variables for the problem
at hand. For the model potential considered in this work
(Sec. \ref{details}), this is not a problem. For more complicated
systems one can identify such CVs in principle through a recently
proposed method \cite{sgoop}.

\section{Simulation details}
\label{details}
We consider a model 2-state potential (Fig. 1(a)) introduced by De Leon
and Berne \cite{deleonberne}, which has been the subject of numerous
similar studies\cite{straubberne_review} in the past. It is given by
\begin{equation}
\label{db_eqn}
V(x,y) = 4y^2(y^2-1)e^{-4.485 x} + 10 (1-e^{-1.95 x})^2 +1
\end{equation}
All units of energy and mass were defined in terms of
 Eq. \ref{db_eqn}. Let $q=(x,y)$ denote the configurational
 coordinates of the system, and $m$ be the mass. All our simulations
 are performed with $m=$ 8 units and temperature $k_B T = 0.1$ units, with Newton's laws of motion integrated per
 Langevin dynamics with a timestep 0.05 units:
\begin{equation}
m\ddot{q} + {dV(q) \over dq} + m \gamma \dot{q} = F(t)
\label{langevin}
\end{equation}
where $\gamma$ is the friction coefficient, $F(t)$ is a Gaussian
random noise with mean zero and correlation function $\langle F(t)
F(t') \rangle = {2m \gamma \over \beta} \delta(t-t') $. There are many
available schemes for implementing Langevin dynamics - we use the one
from Ref. \onlinecite{bussi_langevin}. For unbiased MD as well as
metadynamics, the simulations were performed separately for 11
different values of the friction coefficient $\gamma$ between
$10^{-4}$ and 10. The bias was constructed as a function of the
spatial coordinate $y$. The well-tempered flavor of metadynamics
\cite{wtm} was used, with the so-called biasing factor that controls
the gradual decay of Gaussian amplitude set to 6. An initial Gaussian
height of 0.1 $k_BT $ and width of 0.1 units were used. Two
different biasing frequencies were used to ascertain sensitivity to the
biasing stride - once every 20,000 integration steps and once every
100,000 steps. These will be denoted as the fast and slow deposition
schemes respectively.

\begin{figure}[h]
    \begin{subfigure}{0.45\textwidth}
        \includegraphics[height=2.85in]{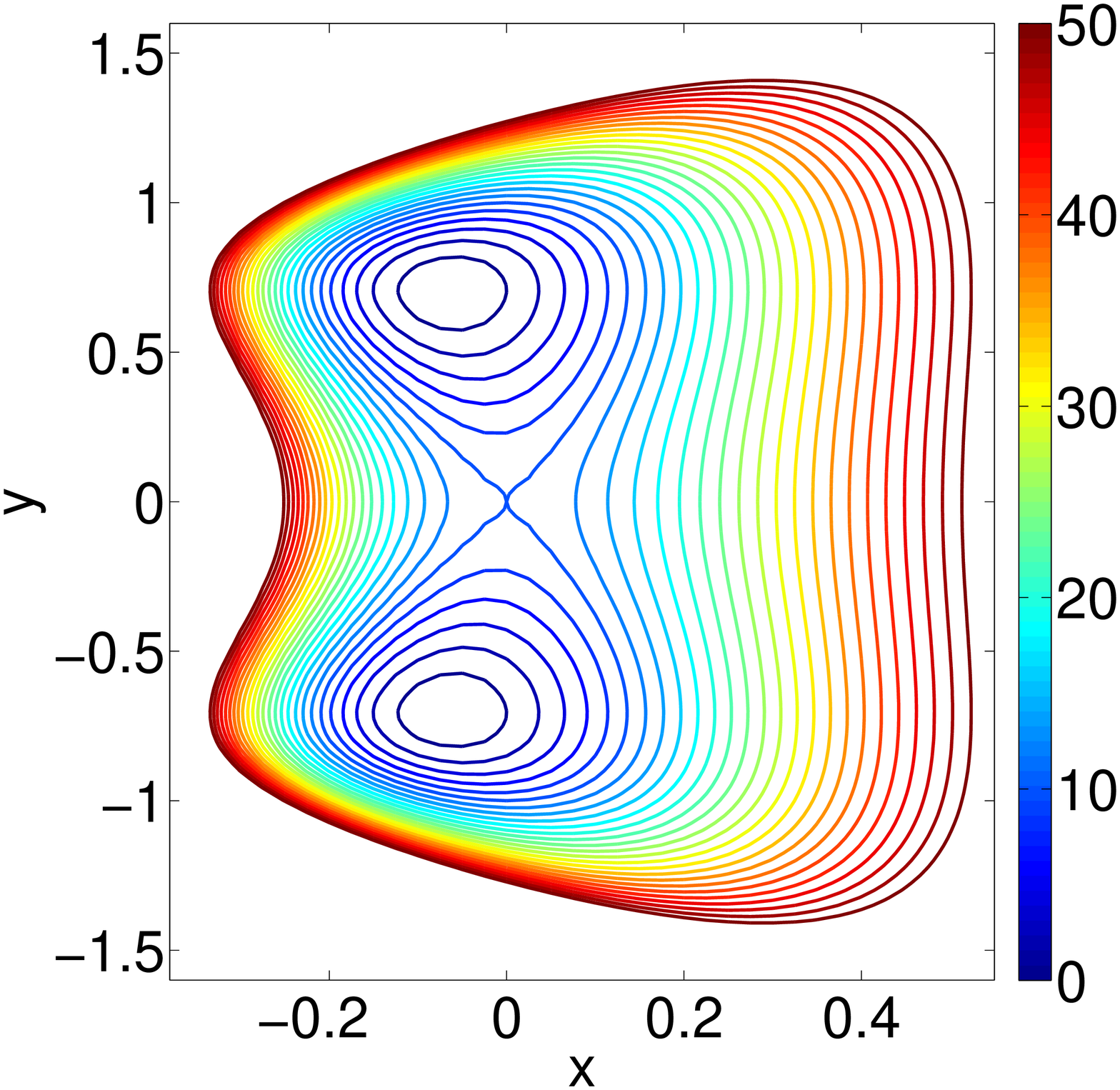}
    \end{subfigure}
    \begin{subfigure}{0.45\textwidth}
        \includegraphics[height=2.85in]{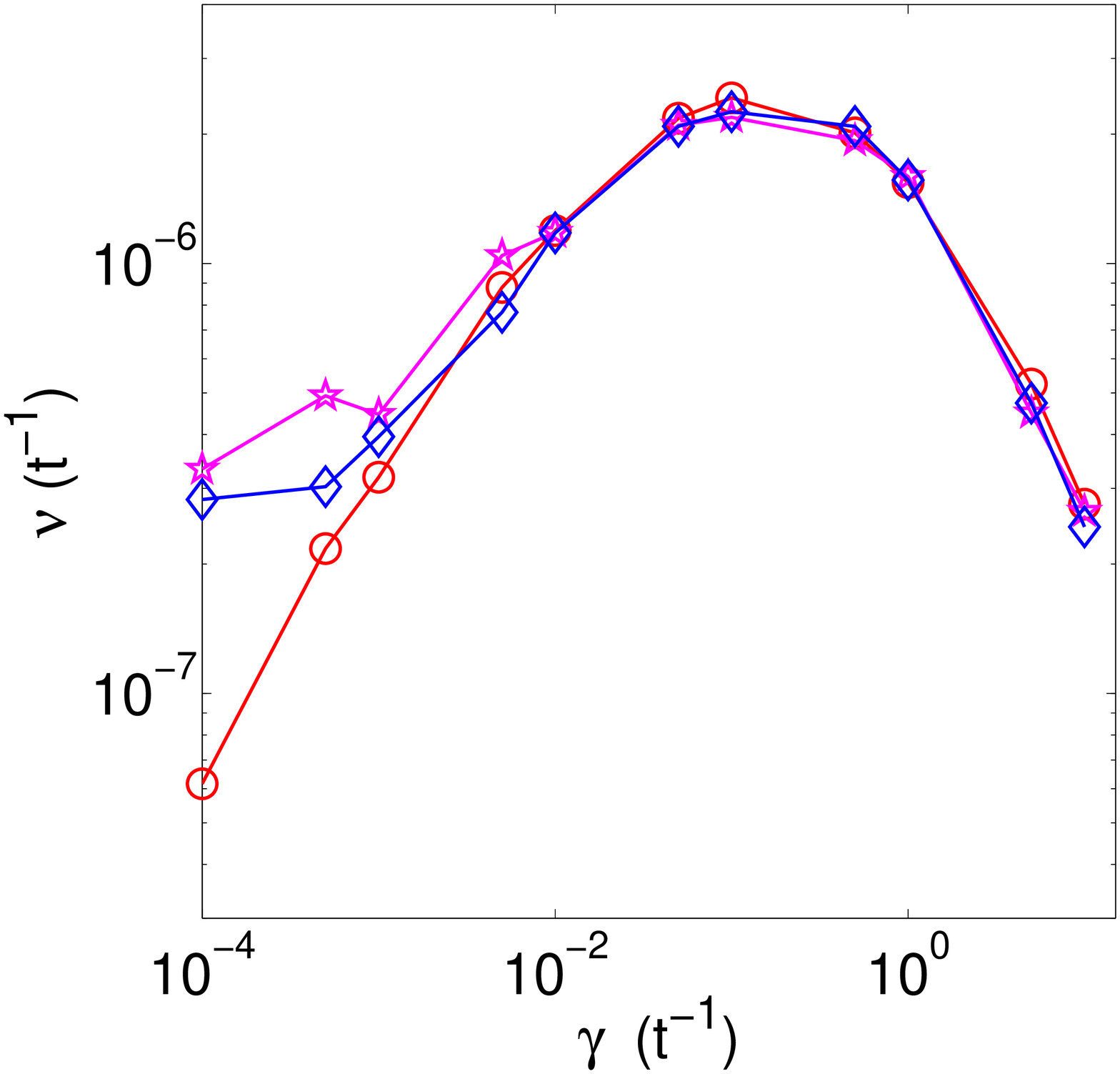}
       \end{subfigure}       
\caption { (a) The 2-state model potential considered in this work, as described in Sec. \ref{details} and in Ref. \onlinecite{deleonberne}. All energies are in $k_B T$ units, with contours spaced every 2 $k_B T$. The barrier here is approximately 10 $k_B T$. (b) The rate $\nu$ of transitions between the two stable states in (a), as a function of the friction coefficient $\gamma$ in Eq. \ref{langevin}. The red circles correspond to unbiased MD estimates. The blue diamonds and magenta stars are for the slow deposition and fast deposition schemes respectively. All units are in accordance with Eq. \ref{db_eqn}.}
\label{fes_noroll}
\end{figure}

\section{Results and discussion}
\label{results}
In Fig. 1(b) we provide the transition rate $\nu$ as a function of the friction coefficient $\gamma$ in Eq. \ref{langevin}, which is kept same for both $x$ and $y$ coordinates. The rate was calculated as  $\nu$ = 1/$\langle t \rangle$ where $\langle t \rangle$ is the average residence time in either of the basins. To filter out spurious recrossing events, we used a minimum residence time criterion of $10^4$ time units, or 2$\times 10^5$ integration steps, to count a transition event as successful. This criterion, which is similar to checking for a plateau  region in the reactive flux formalism \cite{reacflux}, was enforced uniformly across unbiased MD and both the metadynamics schemes. While the absolute magnitude of the rate constant itself shows some sensitivity to the employed minimum residence time cut-off, the conclusions of this work as described below were found to be very robust to this choice. The metadynamics runs with the slow and fast deposition schemes corresponded to acceleration as per Eq. \ref{acceleration} relative to unbiased MD, of approximately 35 and 45 respectively. Naturally, the simulation lengths for the unbiased MD runs were thus correspondingly much longer. The following salient features can be seen from Fig. 1 (b): 
\begin{enumerate}
\item
Both the deposition schemes for metadynamics agree with unbiased MD in the magnitude of the friction $\gamma$ at which turnover in rate occurs, and the dynamics switches from energy diffusion to spatial diffusion.
\item
All three schemes viz. unbiased MD, metadynamics with slow bias deposition and with fast bias deposition agree very well across the spatial diffusion regime, which is not entirely surprising given that the bias was constructed explicitly as a function of the spatial coordinate $y$. 
\item What is more surprising is that MD and metadynamics continue to agree to some extent even in the energy diffusion regime (i.e. low friction $\gamma$). As the friction $\gamma$ is continued to reduce, at some point the rate curves from both the fast and slow deposition schemes cleave off the curve from unbiased MD. The cleavage occurs earlier (that is at higher friction) for the fast deposition scheme than for the slow deposition scheme. This is because the latter is less likely to corrupt dynamical trajectories that take a long time to commit to either of the two stable states even when they have crossed the barrier.
\end{enumerate}

Thus, to summarize, in this short communication we demonstrate through a numerical example that infrequent metadynamics performs remarkably well in obtaining the rate constant for various environmental coupling strengths, especially in the moderate to high friction regime, which is the regime of interest for most solvated biomolecular systems. At very low frictions, the system essentially undergoes a change in reaction coordinate from spatial to one that quantifies energy transfer between the reaction coordinate and the other intramolecular degrees of freedom (i.e. IVR or intramolecular vibrational relaxation). Yet, to some extent even in this regime we find that metadynamics can give accurate rate constants. This work thus supports the use of the approach developed in Ref. \onlinecite{meta_time} for obtaining kinetic rate constants.

Finally, we would like to suggest that such a test that ascertains the accuracy of rates across dynamics regime change be performed on any method before it is judged to be trustworthy as a tool for enhancing molecular dynamics and obtaining rate constants. \\

\textbf{ACKNOWLEDGMENTS}\\

This work was supported by
 grants from the National Institutes of Health [NIH-GM4330] and the 
Extreme Science and Engineering Discovery Environment (XSEDE) [TG-MCA08X002].

	\bibliography{kramers}
	\end{document}